\documentclass[namedreferences]{SolarPhysics}
\usepackage[optionalrh]{spr-sola-addons}

\usepackage{solaheader}
\newcommand{\solareceiveddate}{2 August 2009}
\newcommand{\solaaccepteddate}{27 August 2009}
\newcommand{\solaonlinedate}{3 October 2009}
\newcommand{\doi}{10.1007/s11207-009-9447-1}
\renewcommand{\solavolume}{260}
\renewcommand{\firstpage}{251}
\renewcommand{\lastpage}{260}
\renewcommand{\solayear}{2009}
\usepackage{graphicx}        % For eps figures, newer & more powerful
\usepackage{amssymb}         % useful mathematical symbols
\usepackage{url}             % For breaking URLs easily through lines
\usepackage{epstopdf}
\usepackage{amsmath}
\newcommand{\apj}{\textit{Astrophys. J.}}

\newcommand{\gafd}{\textit{Geophys. Astrophys. Fluid Dyn.}}

\newcommand{\mnras}{\textit{Mon. Not. Roy. Astron. Soc.}}

\begin{document}

\begin{opening}

\title{Nonlinear Evolution of Axisymmetric Twisted Flux Tubes
in the Solar Tachocline}

\author{R. \surname{Hollerbach}$^{1,2}$\sep P.S. \surname{Cally}$^{2}$\sep}
\runningauthor{R. Hollerbach, P.S. Cally}

\runningtitle{Twisted Flux Tubes in the Solar Tachocline}

\institute{$^{1}$
Department of Applied Mathematics, University of Leeds, Leeds LS2 9JT,
United Kingdom\\
(email: \url{rh@maths.leeds.ac.uk})\\
$^{2}$
Centre for Stellar and Planetary Astrophysics, School of Mathematical
Sciences, Monash University, Victoria 3800, Australia\\
(e-mail: \url{paul.cally@sci.monash.edu.au})}

\begin{abstract}
We numerically study the evolution of magnetic fields and fluid
flows in a thin spherical shell.  We take the initial field to be a
latitudinally confined, predominantly toroidal flux tube.  For
purely toroidal, untwisted flux tubes, we recover previously known
radial-shredding instabilities, and show further that in the nonlinear regime
these instabilities can very effectively destroy the original field.  For
twisted flux tubes, including also a poloidal component, there are several
possibilities, including the suppression of the radial-shredding instability,
but also a more directly induced evolution, brought about because twisted flux
tubes in general are not equilibrium solutions of the governing equations.
\end{abstract}

\keywords{Interior, Tachocline; Instabilities; Magnetohydrodynamics}

\end{opening}

\section{Introduction}

The solar tachocline is the transition zone between the uniformly rotating
radiative interior and the differentially rotating convection zone.  First
discovered helioseismically in the early 1990s, it continues to be probed
to this day.  Of particular interest is how thin it is; the angular-velocity
profile changes dramatically over no more than about 5\% of the solar radius.
This intense shear is one of the reasons why the tachocline is now generally
believed to be the seat of the solar dynamo, with poloidal fields being
sheared out to produce very strong toroidal fields.  
See\inlinecite{hrw07}for reviews of many different aspects of the tachocline,
including its role in the solar dynamo.

Given the combination of strong magnetic fields and shear, it was quickly
realized that so-called magneto-shear instabilities could potentially play
an important role in the dynamics of the tachocline.  Indeed, in other
contexts, long before the tachocline was even discovered, it had been known
that both magnetic fields \cite{gough66} and shear \cite{watson81} separately
can lead to instabilities.  The first to study this problem in the tachocline
context were\inlinecite{gilfox97}$\!\!$; one very interesting result that they
obtained was that a combination of magnetic fields and shear can be unstable
even though each ingredient separately would be stable.  \citeauthor{gilfox97}
carried out a 2D calculation, confined to the surface of a sphere, with no
radial variations allowed.  Subsequent work extended this to quasi-3D
shallow-water and 3D thin-shell models, including also both linear onset and
nonlinear equilibration studies.  See\inlinecite{gc07}for a review of this
work.

We will consider a different type of two-dimensionality, namely axisymmetric
solutions.  These have not received as much attention as some of the 3D
solutions, but recent linear-onset calculations \cite{cdg08,dgcm09}
indicate that if the toroidal field is concentrated into latitudinal bands,
instabilities may arise that shred the field in the radial direction.
In this paper we consider the nonlinear evolution of such banded fields.
For purely toroidal, untwisted flux tubes, we obtain results in good
qualitative agreement with the linear-onset calculations.  For mixed toroidal
plus poloidal, twisted flux tubes (which were not considered before) we show
that the field evolves not just {\it via} the onset of instabilities, but much
more directly, simply because in general it is not an equilibrium solution of
the governing equations.  For twisted flux tubes there is then a variety of
possible outcomes, depending on the strengths of both the toroidal and
poloidal components.

\section{Equations} \label{sec:equations}

The equations we wish to solve are the (Boussinesq) Navier-Stokes equation
$$\frac{\partial{\bf v}}{\partial t} + {\bf v\cdot\nabla v} = -\nabla p
  +{\bf(\nabla\times a)\times a} +S{\bf\hat e}_r +\epsilon\nabla^2{\bf v},
  \eqno(1)$$
the magnetic induction equation
$$\frac{\partial{\bf a}}{\partial t}=\nabla\times({\bf v\times a})
  + \epsilon\nabla^2{\bf a},\eqno(2)$$
and the entropy equation
$$\frac{\partial S}{\partial t} + {\bf v}\cdot\nabla S =
  -N^2{\bf v\cdot\hat e}_r + \epsilon\nabla^2 S.\eqno(3)$$
Here $\bf v$ and $\bf a$ denote the fluid and Alfv\'en velocities,
respectively, and $S$ the entropy.  $N$ is the Brunt-V\"ais\"al\"a frequency,
assumed constant.  Length has been scaled by the inner edge of the tachocline,
so we will be interested in solving this system in the interval $r\in[1,1.05]$.
Time has been scaled by the equatorial rotation frequency, $\bf v$ and $\bf a$
as length/time.

Except for the inclusion of the diffusive terms $(\epsilon\nabla^2\cdot)$, these
equations are the same as in\inlinecite{cdg08}$\!\!$, hereafter referred to as
CDG08.  Some diffusivity must be included here to ensure numerical stability,
but values in the range $\epsilon=10^{-5}$ to $10^{-6}$ yielded similar results,
indicating that diffusivity is not significantly affecting the evolution.  The
results presented here are all at $\epsilon=2\times10^{-6}$.

While the equations may be much the same, the subsequent analysis is very
different from that of CDG08.  They considered only the linear onset of
instability, and only for high radial wavenumber $(k)$, thereby ultimately
eliminating $r$ entirely, instead simply having $k$ as a parameter in the
equations.  In contrast, we are interested in a direct numerical solution, in
the finite interval $r\in[1,1.05]$, allowing an arbitrary radial dependence,
and including also the full nonlinear evolution of the solutions.

For axisymmetric solutions, it is convenient to decompose $\bf v$ and $\bf a$
as
$${\bf v}=v{\bf\hat e}_\phi + \nabla\times(\psi{\bf\hat e}_\phi),\qquad
  {\bf a}=b{\bf\hat e}_\phi + \nabla\times(a{\bf\hat e}_\phi).\eqno(4)$$
The toroidal parts, $v$ and $b$, are the azimuthal components of the given
vectors; the poloidal parts, $\psi$ and $a$, are the streamfunctions of the
meridional components.  The original Equations (1) and (2) then become
$$\frac{\partial}{\partial t}v=P_1(b,a) - P_1(v,\psi) + \epsilon D^2v,
       \eqno(5)$$
$$\frac{\partial}{\partial t}D^2\psi=
  \frac{1}{r}\frac{\partial}{\partial\theta}S + P_2(b,b) + P_2(D^2a,a)
   -P_2(v,v)-P_2(D^2\psi,\psi) + \epsilon D^4\psi,\eqno(6)$$
$$\frac{\partial}{\partial t}b=P_2(v,a) - P_2(b,\psi) + \epsilon D^2b,
       \eqno(7)$$
$$\frac{\partial}{\partial t}a=P_1(\psi,a) + \epsilon D^2a,\eqno(8)$$
where
$$D^2=\nabla^2 - 1/(r\sin\theta)^2,\eqno(9)$$
and
$$P_1(X,Y)={\bf\hat e}_\phi\cdot
  \bigl[\bigl(\nabla\times(X{\bf\hat e}_\phi)\bigr)
  \times\bigl(\nabla\times(Y{\bf\hat e}_\phi)\bigr)\bigr],\eqno(10)$$
$$P_2(X,Y)={\bf\hat e}_\phi\cdot\nabla\times
  \bigl[(X{\bf\hat e}_\phi)\
  \times\bigl(\nabla\times(Y{\bf\hat e}_\phi)\bigr)\bigr].\eqno(11)$$

We then wish to solve (5) and (6), with stress-free boundary
conditions
$$\frac{\partial}{\partial r}\Bigl(\frac{v}{r}\Bigr)=\psi=
  \frac{\partial^2}{\partial r^2}\psi=0\qquad
  {\rm at}\quad r=1,1.05,\eqno(12)$$
(7) and (8), with perfectly conducting boundary conditions
$$\frac{\partial}{\partial r}\bigl(br\bigr)=a=0\qquad
  {\rm at}\quad r=1,1.05,\eqno(13)$$
and (3), with $S=0$ at $r=1,1.05$.  These boundary conditions are somewhat
artificial, but {\it any} boundary conditions would necessarily be artificial.
To properly capture all of the dynamics of the tachocline, it should not be
studied in isolation, but rather as part of a global model that also includes
the interior and the convection zone.  However, while there are models that
aim in this direction \cite{ASH}, they are so complicated that one cannot
focus specifically on aspects such as tachocline instabilities.  To study
tachocline instabilities, one must therefore adopt simplified models such as
ours, despite the inevitably unnatural boundaries where the real Sun has none.

We solve these equations using the numerical code described
by\inlinecite{h2000}$\!\!$, in which the radial structure is expanded in terms
of Chebyshev polynomials, the angular structure in terms of spherical harmonics,
and the time-stepping is done {\it via} a second-order Runge-Kutta method.
Resolutions ranging from $50\times1600$ to $80\times2400$ in $(r,\theta)$ were
used, and were all checked to ensure that the results were adequately resolved.
A timestep of $10^{-3}$ was sufficiently small to ensure stability.

To facilitate comparison with CDG08, we choose initial conditions much the
same as theirs.  The initial entropy is simply $S=0$, so any buoyancy effects
arise entirely out of the subsequent evolution.  For the flow, we take the
solar-like differential rotation profile $\omega=v/(r\sin\theta)=1-0.18
\cos^2\theta$, where we note that (1) is written in a non-rotating frame,
so $\omega$ here must include an overall rotation, not just the
Pole-to-Equator differential rotation $-0.18\cos^2\theta$.

For the toroidal field, we take
$$b=Ap\bigl[{\mathrm e}^{-4(\mu-d)^2/W^2(1-d^2)}
           -{\mathrm e}^{-4(\mu+d)^2/W^2(1-d^2)}\bigr]\sin\theta/r,\eqno(14)$$
where $\mu=\cos\theta$.  That is, $b$ consists of two oppositely directed
bands in the two hemispheres, with position $d$, latitudinal bandwidth $W$,
and amplitude $A$.  The normalization factor $p$ is adjusted such that the
maximum field strength $\max_\theta(b)$ is $A/2$.

Values of $A$ up to one, corresponding in dimensional terms to $\max_\theta(b)
\approx10^5$ G, are of greatest interest in the solar context, but other
(especially younger) solar-type stars may well have even larger values.  We
will present runs for $A=1$ and $2$.  A range of possibilities for $d$ and $W$
were considered, and yielded qualitatively similar results.  We therefore show
results only for $d=0.5$, corresponding to bands situated $\pm30^o$ from the
Equator, and $W=\pi/36$, corresponding to a bandwidth of $5^o$.

Thus far our initial conditions are exactly as in CDG08; see
also\inlinecite{dg99}$\!\!$, who were the first to introduce banded toroidal
fields of this type.  To this toroidal field, we now add the poloidal field
$$a=A'p'\bigl[{\mathrm e}^{-4(\mu-d)^2/W^2(1-d^2)}
             +{\mathrm e}^{-4(\mu+d)^2/W^2(1-d^2)}\bigr]\sin\theta(r-1)(r-1.05).
         \eqno(15)$$
The poloidal field thus has the same banded structure as the toroidal, but $a$
is equatorially symmetric, whereas $b$ is anti-symmetric.  Both of these
symmetries correspond to the standard ``dipole'' solutions of solar dynamo
theory.  The differing radial dependencies, $(r-1)(r-1.05)$ for $a$ {\it versus}
$1/r$ for $b$, are dictated by the different boundary conditions (13) that $a$
and $b$ are supposed to satisfy.  The normalization factor $p'$ is adjusted
such that $\max_\theta(a)$ is $A'$.

The amplitude $A'=fA$, so the factor $f$ gives the ratio of poloidal to
toroidal fields.  In addition to the $f=0$, untwisted flux tubes, we will
take $f=\pm10^{-4}$ and $\pm4\times10^{-4}$ for the twisted flux tubes.
The sign of $f$ determines whether the tubes are twisted in a left- or
right-handed sense.  It is not certain which is more appropriate to the
tachocline \cite{Fan04}, so we consider both possibilities, and show that they
yield qualitatively similar behavior.  Regarding the amplitudes of $f$, for
$A=1$ and $|f|=10^{-4}$, the maximum values of $(a_r,a_\theta,a_\phi)$ are
$(0.0022,0.0092,0.5)$.  Even for $|f|=4\times10^{-4}$ the field is thus
predominantly azimuthal, with only around one twist over the full circumference
$\phi\in[0,2\pi]$.  We will see though that even such relatively weak poloidal
fields as this can significantly influence the subsequent evolution.

\section{Results} \label{sec:results}

\subsection{Untwisted Flux Tubes} \label{subsec:untwist}

\begin{figure}
\includegraphics[scale=1.0]{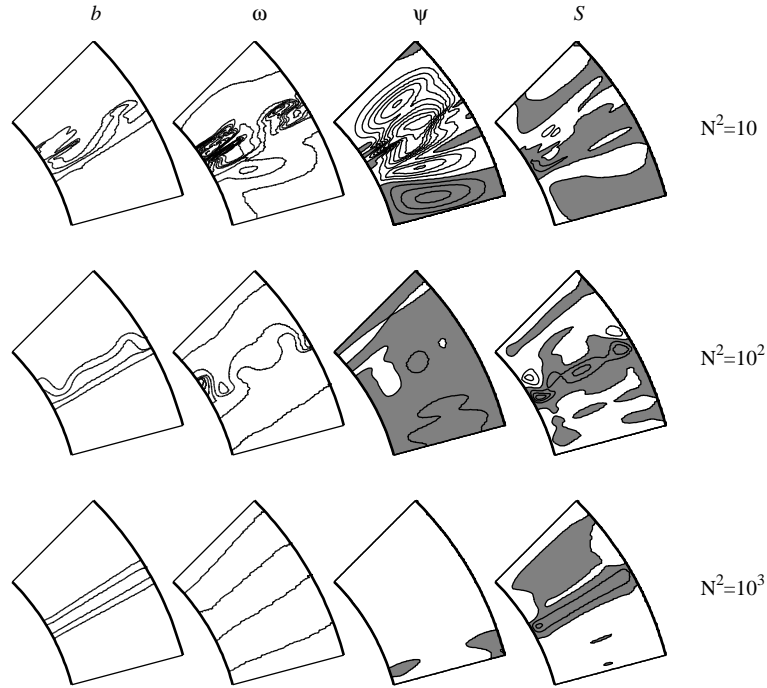}
\caption{From left to right, contours of $b$, $\omega$, $\psi$, and $S$,
with contour intervals of 0.2, 0.02, $5\times10^{-5}$, and 0.1, respectively.
In the plots for $\psi$ and $S$, grey denotes positive values, white negative
($\psi>0$ corresponds to clockwise circulation).  From top to bottom, the three
rows are for $N^2=10, 10^2, 10^3$.  $A=1$ and time $t=5$ for all three.  The
range of $r$ is 1 to 1.05, but has been stretched by a factor of 10, and hence
looks like 1 to 1.5.}
\end{figure}

We begin by considering the influence of the stratification.  Figure 1 shows
results for $N^2=10,10^2,10^3$, all for $A=1$ and $f=0$.  Within each row,
the first panel shows the toroidal field $b$, the second panel the angular
velocity $\omega=v/(r\sin\theta)$, the third the meridional circulation $\psi$,
and the fourth the entropy $S$.  (According to Equation (8), if $a=0$
initially, it will remain zero.)  Only the range $\theta\in[45^o,75^o]$ is
shown, centered on the flux tube at $\theta=60^o$.  The radial direction has
been stretched by a factor of 10, that is, the actual gap $r\in[1,1.05]$ has
been stretched to look like $[1,1.5]$.

\begin{figure}
\includegraphics[scale=1.0]{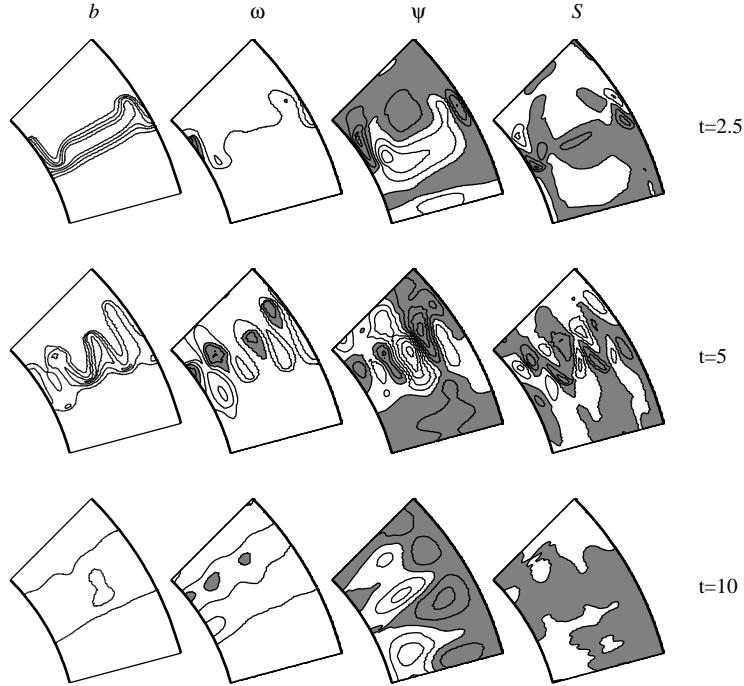}
\caption{From left to right, contours of $b$, $\omega$, $\psi$, and $S$,
with contour intervals of 0.2, 0.05, $5\times10^{-5}$, and 1, respectively.  In
the plots for $\omega$, grey indicates regions where $\omega>1$; for $\psi$
and $S$, grey denotes positive values, as in Figure 1.  $N^2=10^3$, $A=2$,
and from top to bottom $t=2.5$, 5, and 10.}
\end{figure}

\begin{figure}
\includegraphics[scale=1.0]{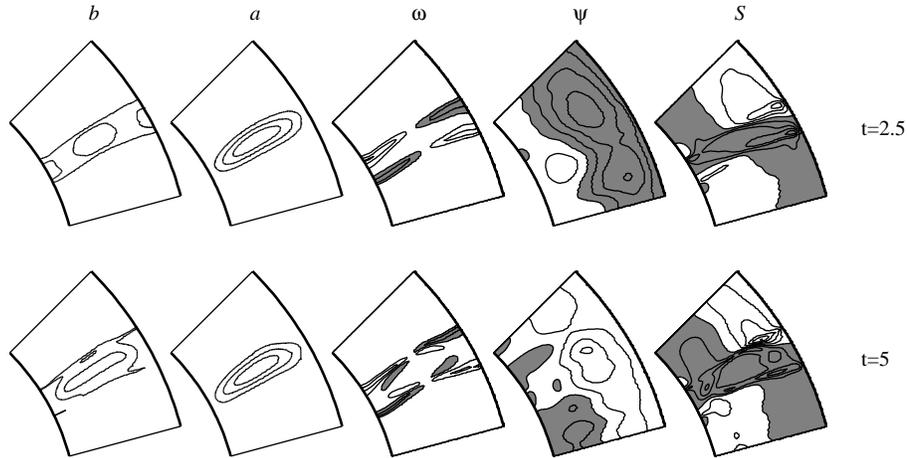}
\caption{From left to right, contours of $b$, $a$, $\omega$, $\psi$, and $S$,
with contour intervals of 0.2, $10^{-4}$, 0.1, $2\times10^{-5}$, and 0.2,
respectively.  In the plots for $\omega$, grey indicates regions where
$\omega>1$; for $\psi$ and $S$, grey denotes positive values, as in Figure 1.
$N^2=10^3$, $A=1$, and $f=4\times10^{-4}$, corresponding to clockwise
circulation for the poloidal field.  The top row is at $t=2.5$, the bottom row
at $t=5$.}
\end{figure}

\begin{figure}
\includegraphics[scale=1.0]{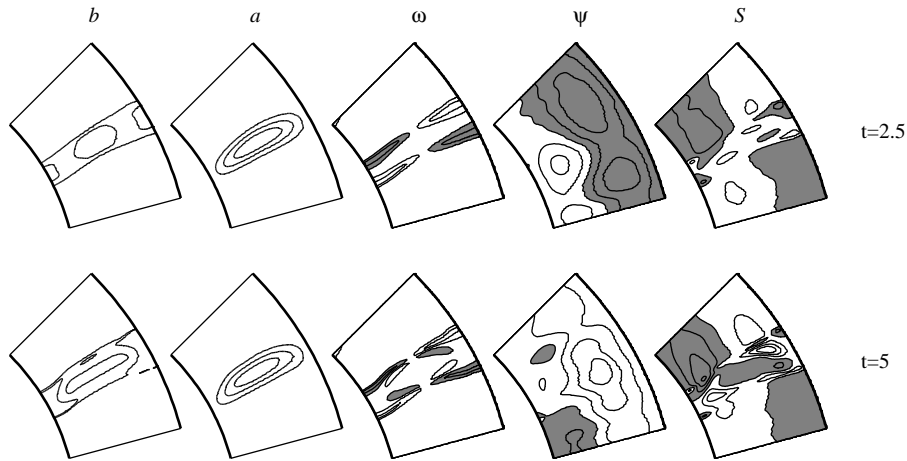}
\caption{From left to right, contours of $b$, $a$, $\omega$, $\psi$, and $S$,
with contour intervals of 0.2, $10^{-4}$, 0.1, $2\times10^{-5}$, and 0.2,
respectively.  $N^2=10^3$, $A=1$, and $f=-4\times10^{-4}$, corresponding to
counter-clockwise circulation for the poloidal field.  (Incidentally, it is
perhaps explicitly worth noting that because of the ten-fold stretching in the
radial direction, $a_\theta$ is actually ten times greater than the spacing of
the contour levels might suggest.  So in fact $a_\theta$ is greater than $a_r$.)}
\end{figure}

Turning to the variation with $N^2$, we see that for $N^2=10$ the
solution has changed significantly from its initial condition, whereas for
$N^2=10^3$ it is almost unchanged.  The reason for this is easy to understand:
If initially only $b$ and $v$ are non-zero -- and if $a$ is always zero --
then according to Equations (5) and (7), the only way (apart from the very
weak dissipation) for either $b$ or $v$ to change is by inducing a meridional
circulation $\psi$.  Now, according to Equation (6), the terms $P_2(b,b)-
P_2(v,v)$ will indeed induce such a circulation; these terms are zero only if
$\frac{\partial}{\partial z}(b^2-v^2)=0$, which in general is not the case.
However, for increasingly strong stratification, the buoyancy term $(r^{-1}
\frac{\partial S}{\partial\theta})$ can very effectively suppress the tendency
to drive a meridional circulation.  That is, having only $b$ and $v$ non-zero
is not quite an equilibrium solution to the governing equations, but if $N^2$
is sufficiently large, only a very weak circulation $\psi$ will be induced, so
according to Equations (5) and (7), $b$ and $v$ will remain almost unchanged.
This justifies linear stability analyses such as those of CDG08, where a basic
state is simply imposed for $b$ and $v$.  In the remainder of this paper we
will consider only the strongly stratified case $N^2=10^3$.

Figure 2 shows the effect of doubling the field strength, to $A=2$.  Now even
the strong stratification is not enough to stabilize the solution.  Instead,
we see the development of precisely the radial-shredding instabilities
previously studied by CDG08.  Here though we follow the full nonlinear
evolution, and discover that by $t=10$ the instability has almost completely
obliterated the original flux tube.

\begin{figure}
\includegraphics[scale=1.0]{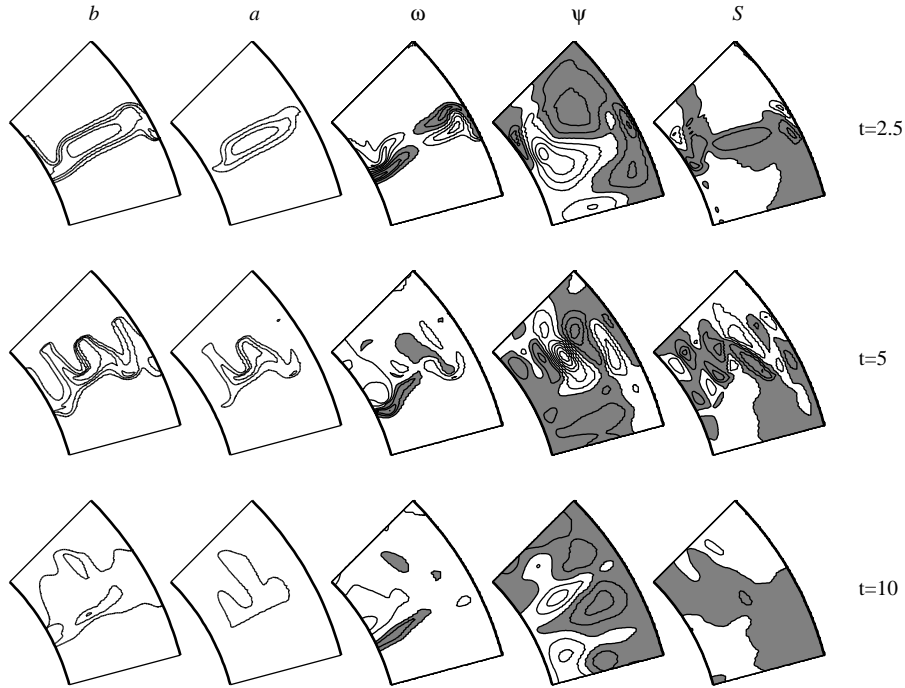}
\caption{From left to right, contours of $b$, $a$, $\omega$, $\psi$, and $S$,
with contour intervals of 0.2, $5\times10^{-5}$, 0.1, $5\times10^{-5}$, and 1,
respectively.  In the plots for $\omega$, grey indicates regions where
$\omega>1$; for $\psi$ and $S$, grey denotes positive values.  $N^2=10^3$,
$A=2$, and $f=10^{-4}$.  From top to bottom $t=2.5$, 5, and 10.}
\end{figure}

\begin{figure}
\includegraphics[scale=1.0]{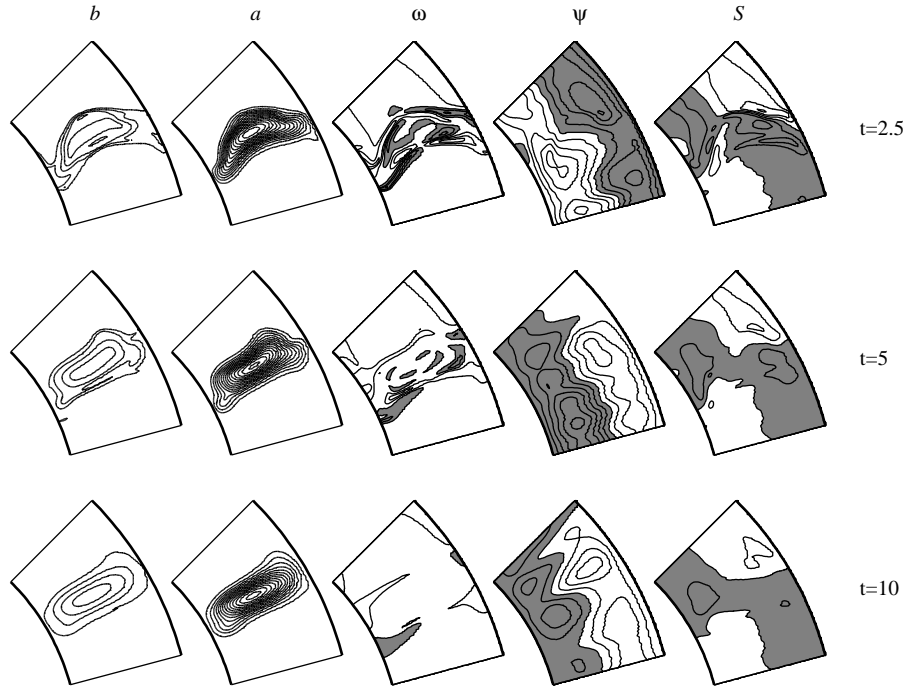}
\caption{From left to right, contours of $b$, $a$, $\omega$, $\psi$, and $S$,
with contour intervals of 0.2, $5\times10^{-5}$, 0.1, $5\times10^{-5}$, and 1,
respectively.  In the plots for $\omega$, grey indicates regions where
$\omega>1$; for $\psi$ and $S$, grey denotes positive values.  $N^2=10^3$,
$A=2$, and $f=4\times10^{-4}$.  From top to bottom $t=2.5$, 5, and 10.}
\end{figure}

\subsection{Twisted Flux Tubes} \label{subsec:twist}

Figures 3 and 4 show the results for $A=1$ and $f=\pm4\times10^{-4}$.
That is, the amplitude $A=1$ is as in Figure 1, too low for the shredding
instability to occur, and indeed it doesn't.  Nevertheless, the solutions also
do not remain virtually stationary, as in the bottom row in Figure 1.  Instead,
we see the development of highly localized jets in the angular velocity $\omega$.
To understand their origin, we return to Equation (5), where the term $P_1(b,a)$
is clearly responsible; this term is zero only if contours of $b$ and $a$
coincide, which in general is not the case.  Furthermore, because there is no
buoyancy force in this equation, no amount of stratification can suppress
this effect, very much unlike Figure 1.

Comparing Figures 3 and 4 in detail, we note that reversing the sign of $a$
(that is, the sense of twist in the tube) has exactly the effect one might
expect; reversing the sign of $P_1(b,a)$ simply reverses the jets.  Otherwise
the evolution is much the same, and in both cases the flux tubes largely
maintain their strength.  Note also that jets as concentrated as this would
not be detectable by helioseismology, so phenomena such as these could
conceivably exist in the real tachocline.

Figure 5 shows results for $A=2$ and $f=10^{-4}$.  The toroidal field is
therefore as in Figure 2, whereas the poloidal field is half as strong as in
Figure 3 (so the nonlinear term $P_1(b,a)$ is just as strong as in Figure 3).
Initially we see the same jets as in Figures 3 and 4 (and reversing the sign
of $f$ again merely reverses the jets).  By $t=5$ the evolution
is dominated by the same shredding instability as in Figure 2, and the final
result is much the same, with the original flux tube largely destroyed.
Poloidal fields as weak as this therefore have relatively little influence.
However, if we increase the poloidal field strength to $f=4\times10^{-4}$, it
does have a very significant influence, as illustrated in Figure 6.  The
radial-shredding instability is now completely suppressed, and the flux tube
persists up to $t=10$ (and beyond).  Note also how the solution has adjusted
itself so that contours of $b$ and $a$ do now largely coincide, and
correspondingly these localized jets are greatly reduced in strength.

\section{Conclusions} \label{sec:conclusion}

We have considered the evolution of flux tubes in a thin spherical shell,
intended to model solar-type tachoclines.  For untwisted tubes we obtain the
same radial-shredding instabilities that have previously been studied in the
linear regime, and show that in the nonlinear regime these instabilities can
very efficiently destroy the original flux tube, simply by shredding it to
sufficiently short length-scales for it to dissipate.

For twisted flux tubes, there are a number of possibilities.  If the toroidal
field is too weak for instabilities to set in, the solution will nevertheless
evolve, {\it via} the formation of differential-rotation jets, driven directly
by the Lorentz forces associated with the twist in the tube.  These jets induce
a certain amount of structure in the flux tubes, but not enough to disrupt it
as the instabilities did.

If the toroidal field is sufficiently strong, and the poloidal field very weak,
the shredding instabilities develop much as before, and simply overwhelm the
jets driven by the twist.  Finally, if the poloidal field is somewhat stronger
-- but still much weaker than the toroidal -- it can suppress the shredding
instability.  Twisted flux tubes can therefore exist at considerably greater
field strengths than untwisted tubes.

Future work will extend this model to 3D and study the interaction of some of
the effects presented here with some of the previously known non-axisymmetric
magneto-shear instabilities \cite{gc07}.

\begin{acknowledgements}
This work was supported by the UK Science and Technology Facilities Council
Grant No. PP/E001092/1.  RH's visit to Australia was supported by a Royal
Society International Travel Grant.
\end{acknowledgements}

\end{document}